# BIO-INSPIRED REQUIREMENTS VARIABILITY MODELING WITHUSE CASE

Esraa Abdel-Ghani and Said Ghoul

Faculty of Information Technology, Research Laboratory on Bio-inspired Software Engineering, Philadelphia University, Amman, Jordan

## ABSTRACT

**Background**.Feature Model (FM) is the most important technique used to manage the variability through products in Software Product Lines (SPLs). Often, the SPLs requirements variability is by using variable use case modelwhich is a real challenge inactual approaches: large gap between their concepts and those of real world leading to bad quality, poor supporting FM, and the variability does not cover all requirements modeling levels.

**Aims**. This paper proposes a bio-inspired use case variability modeling methodology dealing with the above shortages.

**Method**. The methodology is carried out through variable business domain use case meta modeling, variable applications family use case meta modeling, and variable specific application use case generating.

**Results**. This methodology has leaded to integrated solutions to the above challenges: it decreases the gap between computing concepts and real world ones. It supports use case variability modeling by introducing versions and revisions features and related relations. The variability is supported at three meta levels covering business domain, applications family, and specific application requirements.

**Conclusion**. A comparative evaluation with the closest recent works, upon some meaningful criteria in the domain, shows the conceptual and practical great value of the proposed methodology and leads to promising research perspectives.

## FUNDING

Philadelphia University, Research Project: Bio-inspired systems variability modeling project 12/2013

## 1. INTRODUCTION

Variability modeling is becoming recently a basic technique in Software engineering [4, 10, 17, 27].Software Product Line (SPL)represents a set of products belonging to the same family where they have shared features in particular domain. This leads to saving time and effort to build and improve products[9, 12, 18].SPL also presents and regulates the variability through different domain products[2, 11].One of the most important ways to define and manage the variability through products, during SPL design, is feature model[16, 34]. A software Feature Model (FM)is represented as a tree that consists of a set of features and the relationships that connect between them. It is related to software requirements and depends on it[5, 6, 31, 33, 40].

Software requirements deal with stakeholder needs which have to be supported by software. One of the ways to represent the software requirements is by using use case diagrams[1, 26, 37, 42]. A use case diagram represents interactions between user and system components that display the relationships between the user and the use cases in different situations. The requirements modeling is still a challenge because it needs specific model for each system requirements [7, 14]. This can be enhanced by using variability modeling through SPL which allow the representation of a complete business domain requirements in one model such as feature model [30, 40].

Bio-inspired approaches are based on mapping biological features on computing. At present, the use of bio-inspired concepts in various aspects of the computer domains has become widespread and effective[13, 15, 22, 23, 24]. Its introduction in use case variability modeling constitutes the contribution presented in this work.

Software versioning aims to manage and organize the changes and modifications of software and trace the number of software versions in an easier way. Where all of versions have the same basic functionality, but with development and enhancement depending on stakeholder requirements [29, 32, 35].Software revision deals with the changes and modifications of each software version which aims to manage and handle these changes that may occur on each version. A revisions belong to specific version, where it has specific function depending on requirements[8, 25].

Use Case variability modeling is one way to represent software requirements variability [19]. Several researches' works dealt with use case variability modeling in different ways. In [14], the authors introduced an approach to support incremental and evolving configuration in product line use case models to preserve the original parts of specific use case when the changes occur. In [1], the authors presented an approach to extract and generate use case automatically from specification documents that are written by using natural language.The work [26] proposed an approach to create use case scenario from use case pattern and depending on use case goals. The authors in [28] proposed an approach that helps analysts to find and detect repetitions in use case through an automated way. The authors in [36] proposed a use case metamodel.Whereas in [3] the authors applied the variability concept on extend relationship in UML use case model by proposing three types of variability relations.The authors in [20, 21] proposed automated top-down approach that extracts feature model from use case diagram and use case description for specific application to helping the domain analysts.The authors in [39] proposed an approach that includes a transformation description language (TDL) which derives a set of associated use cases from feature model.

Several other works support use case variability modeling. But, the following challenges are not yet levered in any work: large gap between their concepts and those of real world leading to complexity and bad quality, poor supporting FM, and the variability does not cover all application modeling levels: conceptual (business domain), logical (family of applications), and physical (specific application).

This paper proposes a bio-inspired requirements variability modeling methodology carried out through variable business domain use case meta modeling, variable applications family use case meta modeling, and variable specific application use case generating. The bio-inspired basics of this methodology has leaded to integrated solutions to the above shortages: it decreases the gap between computing concepts and real world ones. It supports use case variability modeling by introducing versions and revisions features and related relations. The variability is supported at three meta levels covering business domain, applications family, and specific application.A comparative evaluation with the closest recent works, upon some meaningful criteria in the domain, shows the conceptual and practical value of the proposed methodology and leads to promising research perspectives.

## 2. A BIO-INSPIRED VARIABILITY MODELING IN USE CASE(BVMUC)

This paragraph introduces the main contribution of the paper through a common, simple, and well known software product that manages some kinds of data lists (stack, array, and queue). The reduction of the gap between variable use case concepts and those of the real world, the three variability meta models of use case, and the enhancement of FM with variability concepts will be presented.

### 2.1 Supporting Example

This paragraph presents an introductory example that will be used through the proposed methodology. It deals with a Software list product. Software requirements are represented using FM. List contains different features, such as: structures, behaviors, methods, etc. The features may be implemented in two ways, static or dynamic. So, each feature is defined as static or dynamic. List product model is composed of different models, such as: static array, dynamic array, static stack, dynamic stack, static queue, dynamic queue, etc.Figure 1 shows a FM of software list product [16].

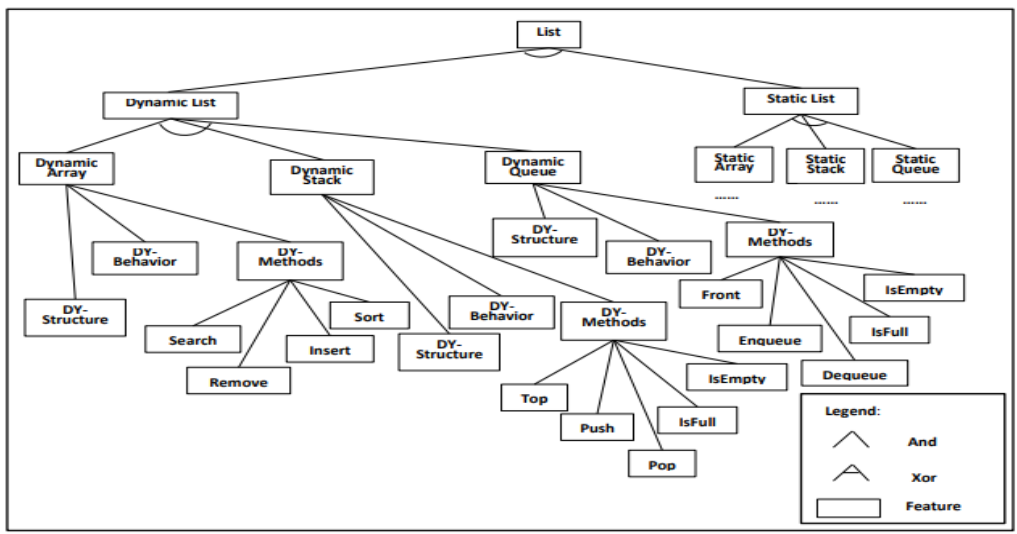

Figure 1:A FM of Software List Product [16].

## 2.2 Proposed Methodology

The first shortage of current approaches supporting use case variability modeling was their distance from real world concepts. This was handled, in this work, by a bio-inspired methodology built on natural Genetics concepts: *Genome*, *Genotype*, and *phenotype* and their supporting processes. Figure 2 displays the methodology three steps, each one carried out by a specific process. The *variable use case feature modeling* process produces variable use case FM from the business domain variable requirements. This process is inspired from the genome concept. The *Configuration feature model processing* process selects the applications family requirements from the variable use case FM and produces variable applications family use case FM. This process is inspired from the genotype concept. The *specific use case generating* process selects an application specific requirement from the application family use case FM and produces a specific use case diagram for that application which might evolve (vary) in the time. This process is inspired from the phenotype concept. Below, the components of the methodology will be developed using EBNF notation [41] which defines the architecture of FM in an abstract way and which is easy to understand.

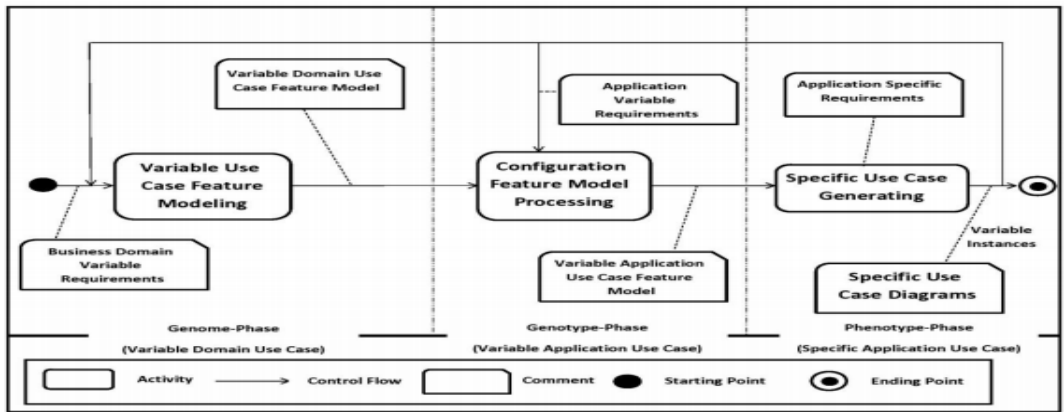

Figure 2:A Bio-Inspired Methodology for Use Case Variability Modeling, using UML notations.

### a. Variable Use Case FM (VUCFM)

In addition to the distance from real world concepts shortage, the introduction of the above bio-inspired methodology has levered the second shortage which is the covering of all requirements modeling levels (business domain, applications family, and specific applications). So, the Variable Domain Use Case FM (VDUCFM), Variable Applications family Use Case FM (VAUCFM), and Specific Use Case FM (SUCFM) are the sub-parts of VUCFM as shown in figure 2. The following formal statement introduces the VUCFM architecture by using EBNF notation:

<VUCFM> = "VUCFM": <Variable Use Case FM name>";"
        <VDUCFM>, <VAUCFM>, <SUCFM>
        "end" "VUCFM"< Variable Use Case FM name >";"

**b. Variable Domain Use Case FM (VDUCFM)**

A VDUCFM is the first (root) variable requirements meta model. It represents the FM of a variable business domain software. Example: the university's *academic management domain* includes students management, teachers management, employees management, etc. It includes a variety of management for different universities requirements. Each one can be managed in a different way. *A variable applications family* might be students' management with all its different requirements (according to different universities). A *specific application* might be students' management at Philadelphia University with its specific requirements. It is generated from its variable applications family(Students management) by fixing some parameters. Each use case can have one or more versions and each version can have one or more revisions depending on the changes in requirements and this reflects the variability concept. Each revision realizes a use case model which includes the use cases, actors and relations. The use case interacts with actors which can Input, Output, or both. Also, use cases can be related with each other using Includes, Extends, or Is-a relation. A VDUCFM contains all variable use case features that can be used to model variable use case according to the domain requirements:

*<VDUCFM>= "VDUCFM" : < Variable Domain Use Case FM name> ";"*
*< Variable Domain Use Case features>*
 *"end" "VDUCFM":< Variable Domain Use Case FM name>";"*

The features of VDUCFM are defined by Variable Domain Use Case features with their relations:

$$<Variable\ Domain\ Use\ Case\ features> = (<Use\ Case\ features>)^+, (<Actor\ features>)^+, \quad (<relations>)^+ ;$$

Use Case features are used to represent use cases. They consist of version features and their relations:

$$<Use\ Case\ features> = "Use\ Case" : <Use\ Case\ features\ name>; \\ (<Version\ features>, <relations>)^+ ;$$

Version features are used to represent versions of use case that define the same use cases but working in different ways (They have the same semantics but they work differently), such as static stack and dynamic stack, both are stack but each one works completely different of the other one. Each version consists of revision features and their relations:

$$<Version\ features> = "Version" : <Version\ features\ name>; \\ (<Revision\ features>, <relations>)^+ ;$$

Revision features of a version are used to represent revisions of use case versions. The revisions deal with the changes occurring in versions:

$$<Revision\ features> = "Revision" : <Revision\ features\ name>; \\ (<Import\ feature>, <relations>)^+ ;$$

Import feature is used to represent import relation between a revision in use case and its origin (from where it is generated). The semantics of import means that a revision R is based on an imported version or revision V (R is obtained by some update in V):

$$<Import\ feature> = "Import"\ (<feature>)^+;$$

Actor feature is used to represent the actors that interact with use case system:

$$<Actor\ feature> = "Actor"\ (<attributes>,\ <relations>)^+;$$

Attributes are used to indicate use cases name, actors' name, etc:

$$<Attributes> = (<Atrribute\_name> :\ <Attribute\_value>)^+;$$

Relations are defined between features (use case features and actor features):

*<Relations>=and|or| xor| is-a| include| extend |composed by|in-interact|out-interact| inOut-interact;*

Each of the relations above has its specific semantics, as it follows:

- B and C: Feature A is composed of the two features B and C together.
- B x or C: Feature A is composed of one of the two features B or C.
- B or C: B or C: Feature A is composed of one of the two features B or C or both.
- B is-a A: Feature A is parent of child feature B. Feature B inherits features from A.
- B include A: Feature B incorporates behaviour sequence of feature A.
- B extend A: Feature B add incremental behaviour to feature A.
- A composed by B: If feature B is part of feature A.
- Ac1 In-interact A: The actor Ac1 interacts with feature A as input.
- Ac1 Out-interact A: The actor Ac1 interacts with feature A as output.
- Ac1 In Out-interact A: The actor Ac1 interacts with feature A as input and output.

The following figures 3a, 3b, and 3c show a graphical representation of VDUCFM for the list example introduced previously. The whole example is in one figure (Genome) but has been divided for arrangement and illustration purposes.

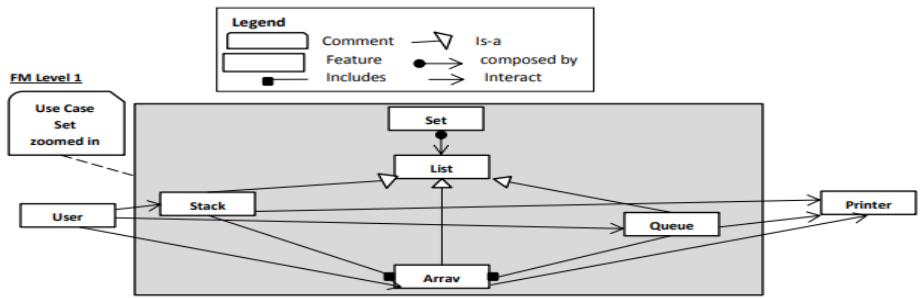

Figure 3a: Variable Domain Use Case Feature Model of Software Set (level 1).

**b. Variable Application Use Case FM (VAUCFM)**

The VAUCFM is the second (middle) variable requirements meta model. It represents variable applications family requirements which are coherent subset of features selected from VDUCFM. It is generated by selecting coherently features that an application family needs from variable domain requirements. The use cases selected should have one or more versions and each version may have one or more revisions. Below an algorithm of Select VAUCFM that defines how a VAUCFM is created from the VDUCFM:

*Select_VAUCFM (VDUC FMi, VAUC FMo, Feature F)*
*{ //this procedure selects a VAUCFM FMo from a given VDUCFM FMi for a feature F*
*        if (F is a revision) then error;*
* else { if (F exist in FMi) then FMo ← generate (tree (F));*
*            //generate is a function generating the FMo from a given FMi for a feature F*
*    else error;*
*            }*
*}*

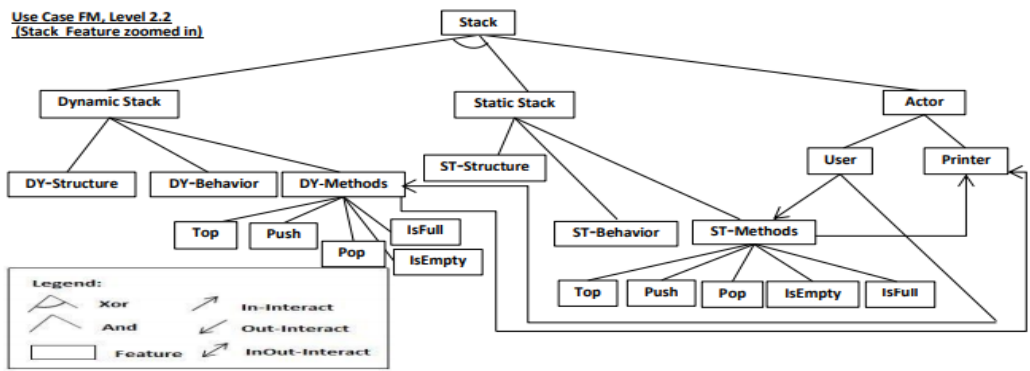

Figure 3b: Example of Variable Domain Use Case FM of Software Set (level 2).

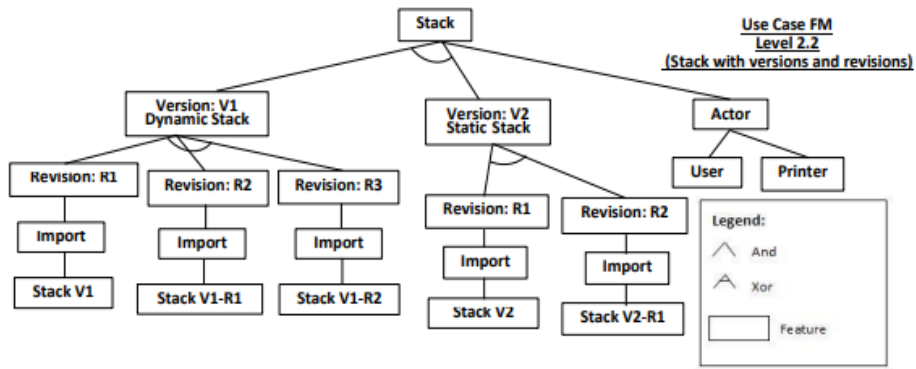

Figure 3c: Example of Variable Domain Use Case FM of Software Set (level 2) with versions and revisions.

In the algorithm above, FMi refers to the input VDUCFM, FMo refers to the output VAUCFM, and F refers to feature characterizing the corresponding applications family. The feature F may be a basic feature, such as Array feature, so it will be selected with all its versions and revisions. Also, it might be a version feature such as Array-Version2 (Static Array), where it will be selected with all its revisions. Never F will be a revision feature such as Array-Version2 (Static Array)-Revision2, because it is not in a variable way to represent variable applications family use case. Also, the FMo will contain all features associated through include, extend, composed by and is-a relations. The following statement shows an example of VAUCFM generation from a VDUCFM:

The Select_VAUCFM operation has SetFM (figure 3) as input and Version2-Stack VAUCFM as output for the static stack (version 2) feature. Figure 4 shows a representation of the obtained VAUCFM.Version2-Stack family that is created according to the previous applications family requirements (Version 2- Stack).

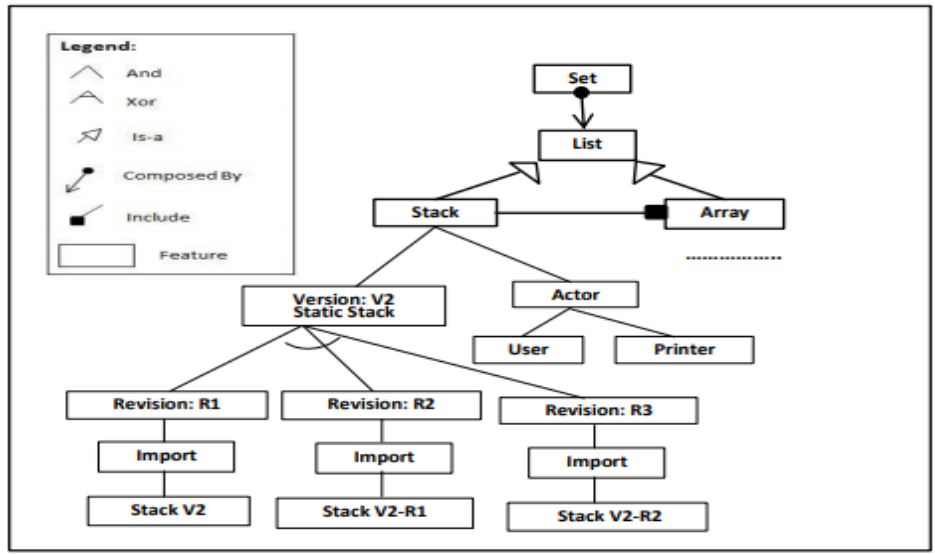

Figure 4: Graphical representation of Static Stack Family FM.

## D. Specific Use Case Fm (Sucfm)

The SUCFM is the third (last) variable requirements meta model. It represents variable application instance of a given VAUCFM for specific application requirements. It is generated from VAUCFM by selecting specific versions and revisions. It might separately evolve and vary through its revisions. In fact, a specific application may generate several revisions from its basic one (selected from its family) depending on its changing needs. At the end, the specific use case FM can be easily converted to UML use case model.

Below an algorithm of select SUCFM that specifies how a SUCFM is created from a VAUCFM:

**Select_SUCFM** (VAUC FMi, SUC FMo, Feature F)
{
      if (F is a revision) then
      {//this procedure selects a SUCFM FMo from a given family VAUCFM FMi for a
      feature F
      if (F exist in FMi) then FMo ← generate (tree (F));
      //generate is a function generating the FMo from a given FMi for a feature F
      else error;

 }
      else error;
}

In the algorithm above, FMi refers to the input VAUCFM, FMo refers to the output SUCFM, and F refers to a feature that will be selected from FMi. F should be revision feature of specific version such as, Array-Version2(Static Array)-Revision2. Also, all features associated with F through include, extend, composed by and is-a relations will be included in FMo.

The following statement shows an example of SUCFM generation from a VAUCFM:

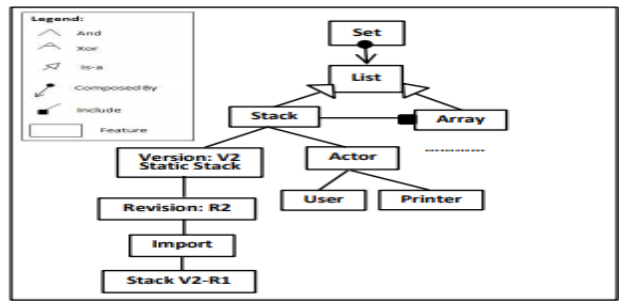

Figure 5: Graphical representation of Revision2 for Static Stack.

Select_SUCFM (StaticStackFM, Revision2-StaticStackFM, Revision2-StaticStack);

The Select_SUCFM operation has Static StackFM family (figure 4) as input and Revision2-StaticStackFM SUCFM as output for the revision 2 of static stack feature. The figure 5 shows a representation of the obtained SUCFM Revision2-StaticStack that is created according to the previous specific application requirements.

The FM represented in figure 5 might be converted to UML use case model by taking the functional use case features that interact with actors features. The revision R2 of version V2 in stack consists of ST structure feature, ST- behaviour feature and ST-methods feature (figure 1). The features that interact with actor is ST-Methods which contain push, pop, is Full, is Empty, and top methods as illustrated by the figure 6 (a) that is generated from figure 3 (b), level 2-2. At the end the figure 6 (b) shows the use case generated from the FM (figure 5).

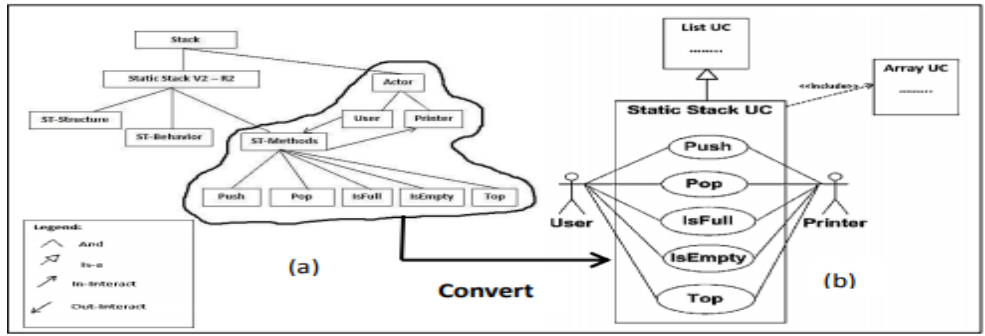

Figure 6: Use case of revision2 static stack.

## 2.3 A Fm Supporting Use Case Variability

The above bio-inspiration has led to enhancing the current FM concepts such that it will be able to support use case variability concepts. New features (like version, revision, import, and actor) and relations (like include, extend, is-a, composed by, in-interact, out-interact, and inOut-interact) were introduced. The new features were explained above and the relations are the common ones of use case modelling.

## 3. RESULTS EVALUATION AND CONCLUSION

### 3.1 Results

The implementation environment of this methodology requires a language that communicates with UML use case and Feature Modelling environments and a database of Variable Domain Use Case FM, Variable Applications family Use Case FM, and Specific Use Case FM with versions and revisions techniques. Software requirements engineering will be more powerful, applicable, and flexible by supporting the Bio inspired Use Case Variability Modelling methodology, since it manages the domain, family, and specific requirements, and the changes that could be occurred. This approach is recommended to be used in any variable software requirements area such as software product lines and domain product lines (the Universities domain systems, Hospitals domain systems, and any other domain systems including many variable applications and functions).

A comparison between the proposed methodology and others recent similar works was developed according to the following meaningful criteria: (1) Distance from real world concepts, (2) Using bio inspired approach for supporting variability in use case modelling, (3) covering all application modelling levels, (4) Supporting version and revision techniques for variability use case modelling, (5) Generating specific use case model from application use case model, (6) Applying integrated and complete methodology, (7) Supporting automated tool, and (8) Improving FM to achieve variability in use case modelling. All the above studied similar works support partially the criteria 5 and 6, but, only some ones support the criteria 7. However, the proposed approach covers all of them.

## 3.2 Conclusion

According to the previous study of use case variability modelling, the current approaches are far from real world concepts. In addition, the current approaches do not deal with variability in all application modelling levels: domain, family, and specific, but they deal with variability in one level. The FM is poor in supporting the use case concepts in a variable way. And, the using of formal languages to formalize the methodologies is still weak, which leads to misunderstanding. So, the variability modelling in use case is still a challenge and needs enhancements.

In this work, a bio-inspired Use Case Variability Modelling was proposed as a methodology to deal with the insufficiency and weakness that was mentioned previously. This methodology aims to generate a specific use case model from variable domain requirements depending on the specific application needs in an efficient way. It is presented through a FM with versions and revisions techniques according to bio inspired concepts to decrease the gap between computing concepts and real world ones. Variable Use Case FM contains three meta models: Variable Domain Use Case Feature presented as Genome, Variable Applications family Use Case FM presented as Genotype, and Specific Use Case FM presented as Phenotype. The versions and revisions techniques are used to handle use cases evolution and maintenance. The bio-inspiration has led to enhancing the FM concepts such that it will be able to support use case variability concepts: new relations (include, extend, is-a, composed by, in-interact, out-interact, and in Out-interact) and new features (version, revision, import, and actor) were introduced. In future work, the bio-inspired variability can be extended to other UML models such as class diagram, sequence diagram, etc. and other models such as ER diagram, flowchart etc.